\documentclass[runningheads]{svmult}

\usepackage{makeidx}   
\usepackage{graphicx}  
\usepackage{subeqnar}  
\usepackage{multicol}  
\usepackage{physprbb}  
\makeindex             

\begin{document}

\title*{Chaotic Quantization:\protect\newline
        Maybe the Lord plays dice, after all?}

\toctitle{Chaotic Quantization:\protect\newline
          Maybe the Lord plays dice, after all?}

\titlerunning{Chaotic Quantization}

\author{Tamas S. Bir\'o\inst{1}
\and Berndt M\"uller\inst{2}
\and Sergei G. Matinyan\inst{3}}

\institute{MTA KFKI RMKI, H-1525 Budapest, P.O.Box 49, Hungary
\and Department of Physics, Duke University, Durham, NC 27708
\and Yerevan Physics Institute, Yerevan, Armenia}


\maketitle

\begin{abstract}
We argue that the {\em quantized} non-Abelian gauge theory can be
obtained as the infrared limit of the corresponding {\em classical}
gauge theory in a higher dimension. We show how the transformation 
from classical to quantum field theory emerges, and calculate Planck's
constant from quantities defined in the underlying classical gauge
theory.
\end{abstract}

\section{Introduction}

The question, how gravitation and quantum mechanics can be merged 
into a consistent unified theory of all fundamental interactions, 
is still open. Logically, either general relativity (GR), or quantum 
mechanics (QM), or possibly both, will have to be replaced by a 
different theory at a more fundamental level. The almost universally 
accepted notion is that it is GR which needs to be replaced, while 
QM presumably provides a truly fundamental description of nature. 
Superstring theory, describing our four-dimensional space-time as 
the low-energy limit of a ten- or eleven-dimensional theory, is 
widely accepted as the most promising approach, but neither the 
precise form nor the full content of this theory is entirely 
understood at the present time.

What about the other option, considering QM as the low-energy limit
of a more fundamental theory? This question has been raised by 't 
Hooft, who conjectured that quantum mechanics can logically arise 
as the low-energy limit of a microscopically deterministic, but 
dissipative theory \cite{tH99,tH01a}. Explicit, but highly simplified 
examples for such a mechanism have been constructed \cite{BJV00,tH01b}.
In a recent publication \cite{BMM01} with Bir\'o and Matinyan, we 
showed how (Euclidean) quantum field theory can emerge in the 
infrared limit of a higher-dimensional, nonlinear classical 
field theory (Yang-Mills theory). We called this phenomenon
{\em chaotic quantization} to distinguish it from the formal
technique named stochastic quantization \cite{PW81}, not
realizing that this term was already introduced by C.\ Beck
several years earlier \cite{Beck95c} for essentially the same 
mechanism. What is special about Yang-Mills fields, however, is
that they ``quantize themselves'', as we shall discuss below.
In Sect.\ 2, we introduce the concept of chaotic quantization of
a system with one degree of freedom, which we extend to field
theory in Sect.\ 3. In Sect.\ 4 we review the chaotic properties
of classical Yang-Mills theory, before we analyze their chaotic
self-quantization in Sect.\ 5. In the final section, we enumerate
and discuss several open problems.

\section{Chaotic quantization}

A classical physical system encodes much more information than
the analogous quantized system. Consider, for instance a point
particle in one dimension. The classical system is defined by
the pair of coordinates $(x,p)$, implying that every point in
the continuous phase space represents a different state. For
the quantum system, on the other hand, the uncertainty relation
limits the localization of the state in phase space to the 
finite element $\Delta x \Delta p \sim \hbar$. This observation
suggests that it may be useful to consider classical systems,
whose internal dynamics results in a self-afflicted loss of 
information. 

Deterministically chaotic systems satisfy this condition. For
such a system, the rate of information loss is encoded in the
Lyapunov exponent $\lambda$, defined as the loarithmic rate of
divergence between neighboring trajectories:
\begin{equation}
|x_1(t)-x_2(t)| \sim \E^{\lambda t} \qquad (\lambda > 0)
\label{eq1}
\end{equation}
or equivalently in the eigenvalue $\gamma$ of the Perron-Frobenius
operator, defined as the logarithmic rate of convergence of the
phase space density to its stationary limit:
\begin{equation}
\rho(x,t) \rightarrow \rho_\mathrm{lim}(x) + \rho'(x)\E^{-\gamma t}
\qquad (\gamma > 0)\, .
\label{eq2}
\end{equation}

Before we pursue this idea further, let us recall
the method of stochastic quantization \cite{PW81}. Consider a
quantum field $\phi(x)$ in Euclidean space with action $S[\phi]$. 
The domain of $\phi$ is formally extended into a fifth dimension
denoted by $\tau$. If the field $\phi(x,\tau)$ obeys the
Langevin-type equation
\begin{equation}
\frac{\partial}{\partial\tau} \phi(x,\tau) =
- \frac{\delta S}{\delta\phi}(x,\tau) + \xi(x,\tau) \, ,
\label{eq3}
\end{equation}
where $\xi(x,\tau)$ represents local white noise defined by
the moments
\begin{equation}
\langle\xi(x,\tau)\rangle = 0 \, ,\qquad
\langle\xi(x,\tau)\xi(x',\tau')\rangle 
= 2 \delta(x-x') \delta(\tau-\tau') \, ,
\label{eq4}
\end{equation}
then the long-time average of any physical observable converges
to the quantum mechanical vacuum expectation value:
\begin{equation}
\lim_{T\to\infty} \frac{1}{T} \int_0^T d\tau\, {\cal O}[\phi(x,\tau)]
= \langle {\cal O}[\phi(x)] \rangle_\mathrm{QM} \, .
\label{eq5}
\end{equation}
Beck's \cite{Beck95a} suggestion was to replace the artificial white 
noise $\xi$ with the ``noise'' generated by a deterministic, 
but chaotic (more precisely: $\varphi$-mixing \cite{Bil78}) process. 

Following Beck \cite{BR87,Beck95b}, let us start by considering a 
dynamical system with two variables $x,y$, which evolves in discrete 
time steps of length $\tau$. We denote the state of the system at 
$t_n=n\tau$ as $(x_n,y_n)$. We are interested in the dynamics in 
the ``physical'' variable $y$, if the motion in $x$ is
chaotic on short time scales. We define the evolution of the
system as follows:
\begin{equation}
(x_{n+1},y_{n+1}) = f(x_n,y_n) 
                  = (T(x_n),\lambda y_n + \tau^{1/2}x_n) \, .
\label{eq6}
\end{equation}
Here the map $T$ is assumed to be $\varphi$-mixing \cite{Bil78}.
Equation (\ref{eq6}) can be considered as the stroboscobic map 
of the differential equation
\begin{equation}
\dot{y} = - \gamma y 
        + \tau^{1/2} \sum_{n=1}^\infty x_{n-1} \delta(t-n\tau) \, ,
\label{eq7}
\end{equation}
with $x_{n+1} = T(x_n)$ and $\lambda = \exp(-\gamma\tau)$.
Obviously, the variables $\{x_n\}$ take the role of the noise 
in this equation.

The Langevin equation (\ref{eq7}) is equivalent to the
Perron-Frobenius equation for the evolution of the phase
space density of an ensemble of systems:
\begin{equation}
\rho_{n+1}(x',y') = \sum_{(x,y)\in f^{-1}(x',y')}
          \frac{\rho_n(x,y)}{\lambda |\partial T/\partial x|}
                  = \sum_{x\in T^{-1}(x')}
          \frac{\rho_n(x,(y'-\tau^{1/2}x)/\lambda)}
               {\lambda |\partial T/\partial x|}
\label{eq8}
\end{equation}
Expanding (\ref{eq8}) into powers of $(\gamma\tau)^{1/2}\equiv\bar\tau^{1/2}$,
taking the limit $\tau\to 0$, and interpolating $\rho_n(x,y)$
to a function $\rho(x,y,t)$ which depends continuously upon time,
\begin{equation}
\rho(x,y,t) = \varphi(x,y,t) + {\bar\tau}^{1/2} a(x,y,t) + {\bar\tau} b(x,y,t)
            + {\bar\tau}^{3/2} c(x,y,t) + \cdots \, ,
\label{eq9}
\end{equation} 
yields a set of coupled equations for the coefficient functions:
\begin{eqnarray}
\varphi(x',y,t) &=& 
     \sum_{x\in T^{-1}(x')} \frac{1}{|\partial T/\partial x|}
                            \varphi(x,y,t)
\nonumber \\
a(x',y,t) &=& 
     \sum_{x\in T^{-1}(x')} \frac{1}{|\partial T/\partial x|}
     \left(a(x,y,t)- x\frac{\partial}{\partial y}\varphi(x,y,t)\right)
\nonumber \\
& \mathrm{etc.} &
\label{eq10}
\end{eqnarray}

Being interested in the dynamics of the physical variable $y$ only,
we define projected functions
\begin{equation}
p_0(y,t) = \int dx\, \varphi(x,y,t) \, ; \qquad
\alpha(y,t) = \int dx\, a(x,y,t) \, ; \qquad
\mathrm{etc.}
\label{eq11}
\end{equation}
For complete maps $T$, it is possible to show that
\begin{equation}
f(x',y,t) = \sum_{x\in T^{-1}(x')} 
            \frac{g(x,y,t)}{|\partial T/\partial x|}
\label{eq12}
\end{equation}
for all $y$ and $t$ implies
\begin{equation}
\int dx\, f(x,y,t) = \int dx\, g(x,y,t) \, .
\label{eq13}
\end{equation}
The first equation in the expansion of the Perron-Frobenius
equation in powers of ${\bar\tau}^{1/2}$ then becomes a tautology, 
while the second one takes the form
\begin{equation}
\frac{\partial}{\partial y} \int dx\, x\, \varphi(x,y,t) = 0 \, .
\label{eq14}
\end{equation}
The desired dynamical equation for the phase space density
$p_0(y,t)$ is obtained as the third equation (at order $\bar\tau$):
\begin{equation}
\frac{\partial}{\partial y} \int dx\, x\, a(x,y,t) = 
\frac{\partial}{\partial y} (y\, p_0(y,t)) +
  {1\over 2}\frac{\partial^2}{\partial y^2}
  \int dx\, x^2\varphi(x,y,t) - \frac{\partial p_0}{\partial t}\, .
\label{eq15}
\end{equation}

If $h(x)$ is an invariant of the map $T$, 
$\varphi(x,y,t) = h(x) p_0(y,t)$ is a solution of the first of
the set of equations (\ref{eq10}) for any function $p_0(y,t)$,
and (\ref{eq14}) requires 
$\langle x\rangle \equiv \int dx\, xh(x) = 0$. Finally, (\ref{eq15})
turns into
\begin{equation}
\frac{\partial}{\partial t} p_0(y,t) = 
\frac{\partial}{\partial y} \left(y\, p_0(y,t)\right) +
\frac{\langle x^2\rangle}{2} \frac{\partial^2}{\partial y^2}p_0(y,t) -
\frac{\partial}{\partial y} \int dx\, x\, a(x,y,t) \, .
\label{eq16}
\end{equation}
The left-hand side and the first two terms on the right-hand side
have the form of a Fokker-Planck (FP) equation; the last term on the
right-hand side represents a source term. For symmetric maps,
$T(x) = T(-x)$, it is easy to see that the source term vanishes,
and the projected phase space density $p_0(y,t)$ satisfies a
homogeneous FP equation. The first correction in ${\bar\tau}^{1/2}$, 
$\alpha(y,t)$, also satisfies a FP equation, but in this case, 
the source term does not vanish. In other words, the deviations 
in the equation for the exact projected phase space density 
$p(y,t) = \int dx\, \rho(x,y,t)$ from a FP equation are 
proportional to ${\bar\tau}^{1/2}$.

An explicit solution can be found for, e.\ g., the Ulam map
$T(x) = 1-2x^2$ with $x\in [-1,1]$. The stationary solution of
the FP equation is:
\begin{equation}
p_0(y,t) \equiv p_0(y) = \left( {2\over\pi} \right)^{1/2} \E^{-2y^2} \, ;
\label{eq17}
\end{equation}
and the full solution written as a power series in ${\bar\tau}^{1/2}$ is:
\begin{equation}
p(y,t) = p_0(y) \left[ 1 + {\bar\tau}^{1/2}\left( 2y - {8\over 3}y^3 +
                O({\bar\tau})\right) \right] \, .
\label{eq18}
\end{equation}

To obtain the Euclidean Schr\"odinger equation, instead of the FP
equation, we need to introduce a potential $V(y)$ and rescale the
auxiliary variable $x$ according to \cite{Beck95c}
\begin{eqnarray}
\hfill T(x) &\longrightarrow & T(x)\, \E^{{\bar\tau} V(y)/\hbar}
\nonumber \\
y_{n+1} = \lambda y_n + {\bar\tau}^{1/2} x_n &\longrightarrow &
y_{n+1} = \lambda y_n 
          + \left(\frac{\hbar{\bar\tau}}{m\sigma^2}\right)^{1/2} x_n \, ,
\label{eq19}
\end{eqnarray}
where $\sigma^2 = \langle x^2\rangle \equiv \int dx\, x^2h(x)$.
Identifing $p_0$ with the Euclidean wavefunction $\psi$, we find that
it satisfies the imaginary-time Schr\"odinger equation:
\begin{equation}
\hbar \frac{\partial}{\partial t} \psi(y,t) +
\left[ - \frac{\hbar^2}{2m} \frac{\partial^2}{\partial y^2}
       + V(y) \right] \psi(y,t) 
\equiv {\cal S}_\mathrm{E} \psi(y,t) = 0 \, .
\label{eq20}
\end{equation}
The correction linear in ${\bar\tau}^{1/2}$ satisfies the same equation,
but with an additional source term:
\begin{equation}
{\cal S}_\mathrm{E} \alpha(y,t) = 
 \left(\frac{\hbar}{2m}\right)^{3/2} \frac{\partial^3}{\partial y^3}
 \psi(y,t) \, .
\label{eq21}
\end{equation}
The complete wavefunction is 
$\tilde\psi = \psi + {\bar\tau}^{1/2}\alpha + \cdots$, which approaches
the Schr\"odinger wavefunction in the limit $\bar\tau=\gamma\tau \to 0$.

The important insight to take away from this derivation
is that an appropriate chaotic process can serve as the source of
the random noise required for the stochastic quantization of a 
dynamical system if the time scale on which the chaotic process 
randomizes is sufficiently short, so that the corrections are
negligible.

\section{Extension to field theories}

Can this mechanism of quantizing classical systems be generalized to
fields $\Phi(x,t)$ with $x$ being a point in three-dimensional space? 
How this can be done is most easily seen, when the field theory is 
defined on a lattice, rather than a spatial continuum. Then all one
needs to do is introduce a map $T$ together with some internal space 
$\{\xi^i\}$ at each lattice point $x_i$. Beck has proposed to define
the evolution law including a nearest neighbor coupling \cite{Beck95c}
\begin{equation}
\xi^i_{n+1} = (1-g) T\left(\xi^i_n\right) + \frac{g}{2d} \sum_{\nu=1}^d 
              \left(\xi^{i+e_\nu}_n + \xi^{i-e_\nu}_n\right) \, ,
\label{eq22}
\end{equation}
which has the continuum limit
\begin{equation}
\xi^{(x)}_{n+1} = T\left(\xi^{(x)}_n\right) + \frac{g'}{2d}
                  \nabla^2 \xi^{(x)}_n \, ,
\label{eq23}
\end{equation}
with appropriate coupling constants $g$, $g'$. We will not follow
this route further here and refer to Beck's recent monograph
\cite{Beck02}.

A more natural approach consists in identifying the local internal 
map space with a compact Lie group ${\cal G}_x$. The simplest 
realization of this idea is the SU(2) gauge theory, i.\ e., the 
Yang-Mills field theory. In this case, the internal degrees of 
freedom (color) of the gauge field provide the local space for
the chaotic map, and the nonlinear dynamics of the gauge field
uniquely defines the map. As we shall see below, there is no need
to introduce new degrees of freedom beyond those provided by the
gauge field (in one additional dimension) itself \cite{BMM01}.
Before exploring this idea in more detail, however, we need to 
review what is known about the chaotic dynamics of Yang-Mills fields.

\section{Interlude: Chaotic properties of Yang-Mills fields}

The chaotic nature of classical non-Abelian gauge theories was first
recognized twenty years ago \cite{MST81,BMM95}. Over the past decade, 
extensive numerical solutions of spatially varying classical non-Abelian 
gauge fields on the lattice have revealed that the gauge field has 
positive Lyapunov exponents that grow linearly with the energy density 
of the field configuration and remain well-defined in the limit of small 
lattice spacing $a$ or weak-coupling \cite{MT92,Gon93}. More recently, 
numerical studies have shown that the $(3+1)$-dimensional classical 
non-Abelian lattice gauge theory exhibits global hyperbolicity.
This conclusion is based on calculations of the complete spectrum 
of Lyapunov exponents \cite{Gon94} and on the long-time statistical 
properties of local fluctualtions of the Kolmogorov-Sinai (KS) entropy 
in the classical SU(2) gauge theory \cite{BMS99}.

It is useful to note some important relationships between ergodic 
and periodic orbits for globally hyperbolic dynamical systems. The 
ergodic Lyapunov exponents $\lambda_{r,i}$ are obtained by 
numerical integration of a randomly chosen ergodic trajectory, 
denoted by its origin $r$. In a Hamiltonian hyperbolic dynamical 
system with $d$ degrees of freedom the sum of its $d-1$ positive 
ergodic Lyapunov exponents is obtained as the ergodic mean of the 
local expansion rate:
\begin{equation}
\lim_{t\rightarrow \infty} h_r(t) \equiv
\lim_{t\rightarrow \infty} {1\over t} \int_0^t \chi(\vec x(t'))\ dt'
=\sum_{i=1}^{d-1}\lambda_{r,i} = h_\mathrm{KS}\ .
\label{eq24}
\end{equation}
Here $h_{\rm KS}$ denotes the Kolomogorov-Sinai entropy and
\begin{equation}
\chi(\vec{x}(t)) = {d\over dt}\ln\det\left( 
{\partial\vec{x}(t) \over \partial\vec x(0)} \right)_{\rm expanding}
\label{eq25}
\end{equation}
is the local rate of expansion along the trajectory $\vec{x}(t)$.
Due to the equidistribution of periodic orbits in phase space it is 
possible to evaluate the ergodic mean in (\ref{eq24}) by weighted sums 
over periodic orbits $p$. In fact, for hyperbolic systems the thermodynamic 
formalism allows to express certain invariant measures on phase space
in terms of averages over periodic orbits \cite{PP90,Gas98}. 
Labeling periodic orbits by $p$ (rather than a starting point), and 
denoting their periods and positive Lyapunov exponents by $T_p$ and 
$\lambda_{p,i}$, respectively, the connection between the positive 
ergodic Lyapunov exponents and those of periodic orbits is given by:
\begin{equation}
h_\mathrm{KS} = \lim_{t\rightarrow\infty}
\frac{\sum_{t\leq T_p\leq t+\varepsilon} \left( \sum_{i=1}^{d-1}
\lambda_{p,i}\right) \exp\left( -\sum_{i=1}^{d-1}\lambda_{p,i}
T_p\right)} {\sum_{t\leq T_p\leq t+\varepsilon}
\exp\left( -\sum_{i=1}^{d-1}\lambda_{p,i}T_p\right)} \ ,
\label{eq26}
\end{equation}
where $\varepsilon>0$ is a small number. The {\em topological pressure} 
$P(\beta)$ is a useful tool for analyzing invariant measures on phase 
space in terms of periodic orbits. This function can be expressed as 
\begin{equation}
P(\beta) = \lim_{t\rightarrow\infty} {1\over t} \ln 
\sum_{t\leq T_p\leq t+\varepsilon} \exp\left( -\beta \sum_{i=1}^{d-1}
\lambda_{p,i}T_p\right)\ .
\label{eq27}
\end{equation}
The relation (\ref{eq26}) then follows from (\ref{eq24}) and from the 
identity $-P'(1)=h_\mathrm{KS}$. 

In order to apply this formalism to the Yang-Mills field, the gauge
theory needs to be formulated as a Hamiltonian system on a spatial
lattice. How to do this is well known: The Hamiltonian lattice gauge
theory was formulated by Kogut and Susskind \cite{KS75} in order to 
study the nonperturbative properties of nonabelian gauge theories, 
such as quark confinement. Denoting the lattice spacing by $a$ and 
the nonabelian coupling constant by $g$, the Hamiltonian for the
SU(2) gauge theory is given by
\begin{equation}
H = \frac{g^2}{2a} \sum_{x,i} \mathrm{tr}\, E_{x,i}^2
  + \frac{2}{g^2a} \sum_{x,ij} (2 - \mathrm{tr}\, U_{x,ij})
\label{eq28}
\end{equation}
Here the $U_{x,i} \in$ SU(2) are called the link variables at point
$x$ in the coordinate direction $i$, and the $E_{x,i} \in$ LSU(2)
denote the color-electric field strengths components. $U_{x,ij}$
denotes the plaquette product of four link variables, starting and
ending at $x$ and circumscribing the elementary square in the $i,j$
directions. The chaotic nature of this theory was demonstrated 
\cite{MT92} by numerical simulations, and Gong \cite{Gon94} obtained 
the complete Lyapunov spectrum for lattice volumes $L^3$ with $L=1,2,3$. 
Bolte et al. \cite{BMS99} extended these calculations to the lattices 
of size $L=4,6$. 

\begin{figure}[htb]
\begin{center}
\includegraphics[width=0.7\linewidth]{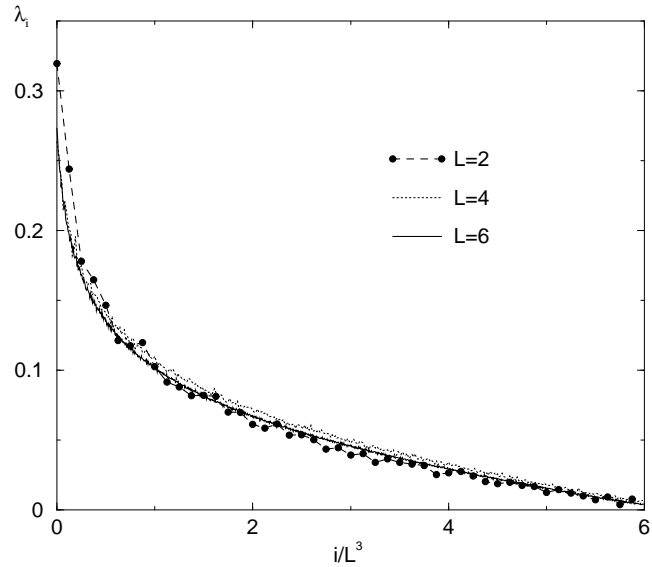}
\end{center}
\caption{Distribution of numerically obtained ergodic Lyapunov
exponents for a classical SU(2) gauge theory on lattices of size
$L=2,4,6$. The index $i$ numbers the Lyapunov exponents and the
abscissa is scaled with $L^3$. The energy per plaquette was
chosen as $1.8/g^2a$.}
\label{fig1}
\end{figure}

For sufficiently long trajectories and fixed energy per lattice site the 
Lyapunov spectrum shows a unique shape, independent of the lattize size,
as shown in Fig.\ref{fig1}. Indeed, for a completely hyperbolic system, 
physical intuition dictates that the KS entropy $h_\mathrm{KS}$ is an 
extensive quantity. For this to be true, the sum over all positive 
Lyapunov exponents must scale like the lattice volume $L^3$ and the shape 
of the distribution of Lyapunov exponents must be independent of $L$. 
Figure \ref{fig1} confirms this expectation.

\begin{figure}[htb]
\begin{center}
\includegraphics[width=0.7\linewidth]{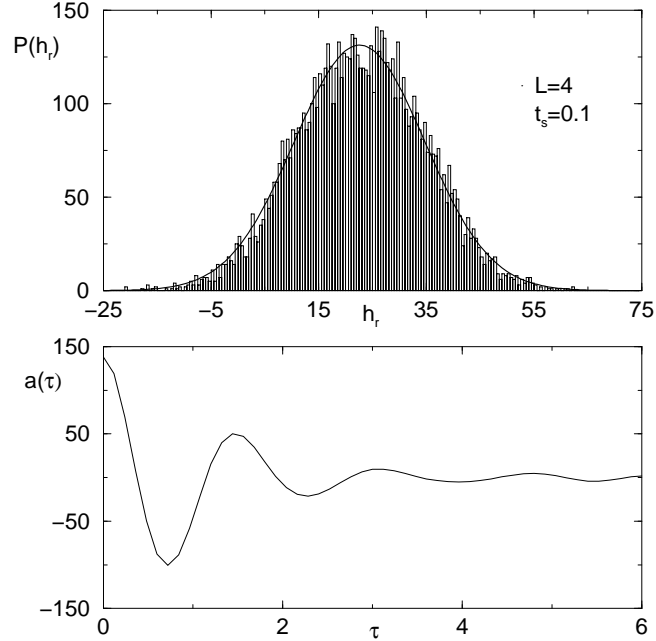}
\end{center}
\caption{Top: The distribution of the sum of local expansion rates 
$h_r(t_\mathrm{s})$ for L=4 and a short sampling time $t_\mathrm{s}=0.1$, 
together with a Gaussian fit. 
Bottom: The autocorrelation function $a(\tau)$ for this distribution.}
\label{fig2}
\end{figure}

The KS entropy is a {\em global} property of the dynamical system. The
next step of detail of the ergodic nature of the system is provided by
the fluctuations in the quasi-local average of $\chi(\vec{x}(t))$.
These fluctuations are obtained by integrating (\ref{eq25}) up to a 
sampling time $t_\mathrm{s}$. For sampling times $t_\mathrm{s}$ much 
longer than the largest correlation time one expects that observables 
sampled along ergodic trajectories exhibit Gaussian fluctuations about 
their ergodic mean. Waddington \cite{Wad96} has shown that for Anosov 
systems (i.e., fully hyperbolic systems on compact phase spaces) the 
probability distribution for $h_r(t_\mathrm{s})$ is Gaussian with mean
$h_\mathrm{KS}$:
\begin{equation}
P[h_r(t_\mathrm{s})] \rightarrow 
\exp\left( - \frac{ (h_r(t_\mathrm{s}) - h_\mathrm{KS})^2 }
{ 2 \Delta h_r(t_\mathrm{s})^2 } \right) \qquad
\mathrm{for}\quad t_\mathrm{s}\rightarrow\infty\ .
\label{eq29}
\end{equation}
and square variance proportional to $P''(1)$:
\begin{equation}
\Delta h_r(t) \rightarrow \sqrt{P''(1)/t} \qquad 
              \mathrm{for}\quad t\rightarrow\infty\ .
\label{eq30}
\end{equation}
The variance (\ref{eq30}) can be related to the autocorrelation function
\begin{equation}
a (\tau) = \langle \chi(\vec{x}(\tau))\,\chi(\vec{x}(0)) \rangle 
- h_\mathrm{KS}^2
\label{eq31}
\end{equation}
of the local ergodic Lyapunov exponents, through the relation
\begin{equation}
t\,(\Delta h_r(t))^2 = 
\int_{-t}^{+t} \left( t-\frac{|\tau|}{t} \right)\, a(\tau)\ d\tau\ 
\rightarrow P''(1)\ .
\label{eq32}
\end{equation}

Figure \ref{fig2} (top) shows the distribution of sampled valued of
$h_r(t_\mathrm{s})$, (obtained on a single, very long trajectory)
for a short sampling interval $t_\mathrm{s} = 0.1$ for a $L=4$ 
lattice. The bottom part of Fig.~\ref{fig2} shows the autocorrelation
function $a(\tau)$ obtained for the same trajectory. A numerical fit
of the function indicates that $a(\tau)$ decays exponentially for
large times. The value predicted for $\Delta h_r$ by 
(\ref{eq30},\ref{eq32}) is in excellent agreement with the value 
obtained by a Gaussian fit to the sampled distribution.

\begin{figure}[htb]
\begin{center}
\includegraphics[width=0.7\linewidth]{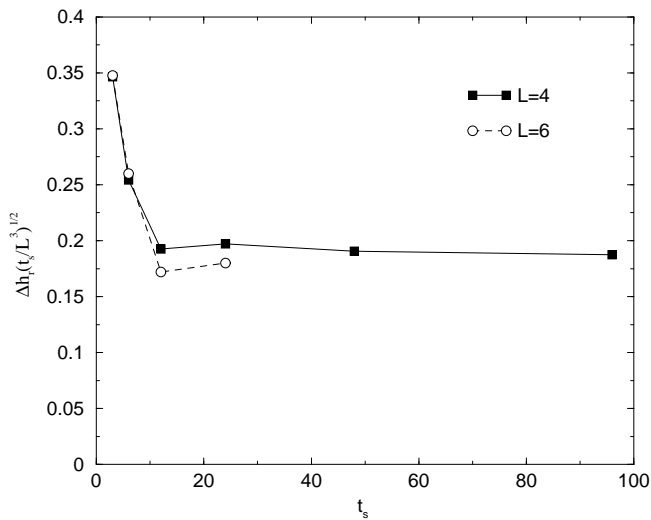}
\end{center}
\caption{$\Delta h_r(t_\mathrm{s})$, scaled with 
$(t_\mathrm{s}/L^3)^{1/2}$, as a function of $t_\mathrm{s}$.}
\label{fig3}
\end{figure}

One can also study how the width of the Gaussian scales with the lattice 
size $L$. To a very good approximation one finds that it is proportional 
to $\sqrt{L^3}$. If one includes the dependence on the sampling time, the 
variance of $h_r$ scales like $\sqrt{L^3/t_\mathrm{s}}$ 
(see Fig.~\ref{fig3}).  As the mean value $h_\mathrm{KS}$ scales like 
$L^3$, this result confirms the Gaussian nature of the fluctuations: 
$\Delta h_r/h_\mathrm{KS} \sim (L^3t_\mathrm{s})^{-1/2}$.

These numerically obtained results provide strong (although not
mathematically conclusive) evidence that the SU(2) lattice gauge
theory is a strongly chaotic (Anosov, $\varphi$-mixing) system with 
properties required for the formalism of Sect. 2.

\section{Yang-Mills fields quantize themselves}

The results discussed in the previous section imply that correlation 
functions of physical observables decay rapidly, and that long-time 
averages of observables for a single initial gauge field configuration 
coincide with their microcanonical phase-space average, up to Gaussian 
fluctuations which vanish in the long observation time limit as 
$t_\mathrm{s}^{-1/2}$. Since the relative fluctuations of extensive 
quantities scale as $L^{-3/2}$, the microcanonical (fixed-energy) average 
can be safely replaced by the canonical average when the spatial volume 
probed by the observable becomes large. In the following we discuss the 
hierarchy of time and length scales on which this transformation occurs
\cite{BMM01}.

According to the cited results, the classical non-Abelian gauge field 
self-thermalizes on a finite time scale $\tau_\mathrm{eq}$ given by the 
ratio of the equilibrium entropy and the KS-entropy, which determines
the growth rate of the course-grained entropy: 
\begin{equation}
\tau_\mathrm{eq} = S_\mathrm{eq}/h_\mathrm{KS}\, .
\label{eq33}
\end{equation}
At weak coupling, the KS-entropy for the $(3+1)$-dimensional
SU(2) gauge theory scales as
\begin{equation}
h_\mathrm{KS} \sim g^2 E \sim g^2 T (L/a)^3\, ,
\label{eq34}
\end{equation}
where $E$ is the total energy of the field configuration and $T$ is
the related temperature related to $E$ by (for SU($N_\mathrm{c}$)
\begin{equation}
\epsilon \equiv \frac{E}{L^2} = 
2(N_\mathrm{c}^2-1) \frac{T}{a^3} + {\cal O}(g^2) \, . 
\label{eq35}
\end{equation}
The equilibrium entropy of the lattice is independent of the energy and 
proportional to the number of degrees of freedom of the lattice: 
$S_\mathrm{eq} \sim (L/a)^3$. The time scale for self-equilibration is 
thus given by
\begin{equation}
\tau_\mathrm{eq} \sim \frac{E}{h_\mathrm{KS} T} \sim (g^2 T)^{-1}\, .
\label{eq36}
\end{equation}
When one is interested only in long-term averages of observables,
it is thus sufficient to consider the {\em thermal} classical gauge
theory on a three-dimensional spatial lattice. Furthermore, on time
scales $t \gg \tau_\mathrm{eq}$, the Yang-Mills field generates a
random Gaussian process, which is required for chaotic quantization.

The dynamic properties of thermal non-Abelian gauge fields at such
long distances have been studied in much detail \cite{Bo98,ASY99,ASY97}.
While these studies have been made exclusively for the thermal quantum 
field theory, their results are readily transcribed to the thermal
classical gauge field with a lattice cutoff. The real-time dynamics 
of the gauge field at long distances and times can be described, at 
leading order in the coupling constant $g$, by a Langevin equation 
\begin{equation}
\sigma_\mathrm{c} {\partial A\over\partial t} = - D \times B + \xi\, ,
\label{eq37}
\end{equation}
where $D$ is the gauge covariant spatial derivative, $B = D \times A$ 
is the magnetic field strength, and $\xi$ denotes Gaussian distributed 
(white) noise with zero mean and variance
\begin{equation}
\langle \xi_i(x,t)\xi_j(x',t') \rangle
= 2\sigma_\mathrm{c} T \delta_{ij}\delta^3(x-x')\delta(t-t')\, .
\label{38}
\end{equation}
Here $\sigma_\mathrm{c}$ denotes the color conductivity \cite{GS93} 
of the thermal gauge field which is determined by the ratio 
$\omega^2/\gamma$ of the plasma frequency $\omega$ and the damping 
rate $\gamma$ of a thermal gauge field excitation. In the classical 
field theory with a lattice cutoff, the color conductivity scales as
\begin{equation}
\sigma_\mathrm{c} \sim 
\left( a \ln[d_{\rm mag}/d_{\rm el}] \right)^{-1} \, .
\label{eq39}
\end{equation}
This relation implies that the color conductivity is an ultraviolet
sensitive quantity, which depends on the lattice cutoff $a$. 

We now consider observers measuring physical quantities on long
time and distance scales ($t,L \gg a, (g^2T)^{-1}$). The random 
process defined by the Langevin equation (\ref{eq37}) generates 
three-dimensional field configurations with a probability 
distribution $P[A]$ determined by the Fokker-Planck equation 
\begin{equation}
\sigma\,\frac{\partial}{\partial t} P[A] \, = \,
\int \! d^3x \, \frac{\delta}{\delta A} \left( T \frac{\delta P}
{\delta A} + \frac{\delta W}{\delta A}\, P[A] \right) \, ,
\label{eq40}
\end{equation}
where $W[A]$ denotes the magnetic energy functional
\begin{equation}
W[A] = \int\! d^3x\, \frac{1}{2} B(x)^2 \, .
\label{eq41}
\end{equation}
Any non-static excitations of the magnetic sector of the gauge field,
i.e. magnetic fields $B(k)$ not satisfying $k\times B=0$, die away 
rapidly on a time scale of order $\sigma_\mathrm{c}/k^2$, where $k$ 
denotes the wave vector of the field excitation. Long-term averages
are determined by the static magnetic field sector weighted by the 
stationary solution of the FP equation (\ref{eq40}):
\begin{equation}
P_0[A] = e^{-W[A]/T}\, .
\label{eq42}
\end{equation}
The observer measures
\begin{equation}
\langle {\cal O}[A] \rangle =
\int {\cal D}A {\cal O}[A] \E^{-W[A]/T} \, .
\label{eq43}
\end{equation}
The magnetic field
\begin{equation}
B_i = \frac{1}{2} \epsilon_{ijk}F^{jk}
\equiv \frac{\sqrt{a}}{2} \epsilon_{ijk}f^{jk}
\label{eq44}
\end{equation}
defines a three dimensional field strength tensor $f^{jk}$, and $W/T$ 
can be identified with the three-dimensional action $S_3$ measured in 
units of Planck's constant  $\hbar = aT$ \cite{BMM01}:
\begin{equation}
\frac{W}{T} \equiv \frac{S_3}{\hbar} =
- \frac{1}{4\hbar} \int\! dx_3 \int\! d^2x\, f^{ik}f_{ik}\, .
\label{eq45}
\end{equation}
The rescaling of the gauge field strength by the fundamental length 
scale $a$ is required for dimensional reasons. The same rescaling 
also determines the three-dimensional coupling constant to be
\begin{equation}
g_3^2 = \frac{g^2}{a} = \frac{g^2T}{\hbar_3} \, .
\label{eq46}
\end{equation}

The central result of this section is that the highly excited classical
$(3+1)$-dimensional Yang-Mills theory reduces to a {\em vacuum}
quantum Yang-Mills theory in three Euclidean dimensions for an 
observer who is only interested in physics at long distance and 
time scales, with vacuum expectation values of the standard form:
\begin{equation}
\langle {\cal O}[A] \rangle_3 =
\int {\cal D}A {\cal O}[A] \E^{-S_3[A]/\hbar} \, .
\label{eq47}
\end{equation}
Planck's constant is determined by two microscopic quantities of
the ``fundamental'' theory, $a$ and $T$:
\begin{equation}
\hbar = a T \, .
\label{eq48}
\end{equation}
The existence of both, a fundamental length scale and an energy
scale, is critical to the emergence of a constant with the
dimensions of an action. 

Some noteworthy comments:
\begin{itemize}
\item
It is important to note that the dimensional reduction is not induced 
by a compactification of the time coordinate, either in real or 
imaginary time. We have not asumed a thermal ensemble of gauge fields 
in the original Minkowski space theory, and the random solution of the 
$(3+1)$-dimensional classical field theory does not satisfy 
periodic boundary conditions in imaginary time. The effective 
dimensional reduction is not caused by a discreteness of the 
excitations with respect to the time-like dimension, but by the 
dissipative nature of the $(3+1)$-dimensional dynamics. Magnetic 
field configurations satisfying $D\times B=0$ can be thought of 
as low-dimensional attractors of the dissipative motion, and the 
chaotic dynamical fluctuations of the gauge field around the 
attractor can be consistently interpreted as quantum fluctuations 
of a vacuum gauge field in 3-dimensional Euclidean space. 
\item
The dimensional reduction by chaotic fluctuations and dissipation 
does not occur in scalar field theories, because there is no 
dynamical sector that survives long-time averaging. Quasi-thermal 
fluctuations generate a dynamical mass for the scalar field 
and thus eliminate any arbitraily slow field modes. An exception 
may be the case where the excitation energy of the scalar field 
is just right to put the quasi-thermal field at the critical 
temperature of a second-order phase transition, where arbitraily
slow modes exist as fluctuations of the order parameter. In the 
case of gauge fields, the transverse magnetic sector is protected 
by the gauge symmetry, and it is this sector which survives the
time average without any need for fine-tuning of the microscopic
theory.
\item
Generalizing the results of Sect.\ 2 and 4, we expect corrections 
to the vacuum quantum field theory in three dimensions to be of 
the order of the relative fluctuations of the KS entropy within a
spatial region of size $\Delta x$:
\begin{equation}
\frac{\Delta h_r}{h_\mathrm{KS}} \sim (g^2 T \Delta x)^{-2}
\sim \left( \frac{a}{g^2 \hbar \Delta x} \right)^2 \, .
\label{eq49}
\end{equation}
If, e.g., $a$ is of the order of the Planck length and $\Delta x$ 
is any physically accessible length scale, the corrections to the
dynamics of the quantized Yang-Mills field will be exceedingly
small, indeed.
\end{itemize}

\section{Open problems}

Our example for the chaotic quantization of a three-dimensional 
gauge theory in Euclidean space raises a number of questions:
\begin{enumerate}
\item
Does the principle of chaotic quantization generalize to higher
dimensions, in particular, to quantization in four dimensions?
\item
Can the method be extended to describe field quantization in 
Minkowski space?
\item
Can gravity be included in this framework? Does the nonlinearity
of classical general relativity provide a source of random noise
at short distances, allowing for an effective quantum theory to
emerge on long distance and time scales?
\end{enumerate}

The first question is most easily addressed. As long as globally
hyperbolic classical field theories can be identified in higher
dimensions, our proposed mechanism should apply. Although we do
not know of any systematic study of dicretized field theories in
higher dimensions, a plausibility argument can be made that
Yang-Mills fields exhibit chaos in (4+1) dimensions. For this
purpose, we consider the infrared limit of a spatially constant
gauge potential \cite{MST81,BMM95}. For the SU($N_\mathrm{c}$) 
gauge field in $(D+1)$ dimensions in the $A_0=0$ gauge, there are 
$3(N_\mathrm{c}^2-1)$ interacting components of the vector potential 
and $3(N_\mathrm{c}^2-1)$ canonically conjugate momenta, which depend 
only on the time coordinate. The remaining gauge transformations and 
Gauss' law allow to eliminate $2(N_\mathrm{c}^2-1)$ degrees of freedom. 
Next, rotational invariance in $D$ dimensions permits to reduce the 
number of dynamical degrees of freedom by twice the number of 
generators of the group SO($D$), i.e. by $D(D-1)$. This leaves a 
$(D-1)(2N_\mathrm{c}^2-2-D)$-dimensional phase space of the dynamical 
degrees of freedom and their conjugate momenta. For the dynamics to be 
chaotic, this number must be at least three. For the simplest gauge 
group SU(2), this condition permits infrared chaos in $2\le D\le 5$ 
dimensions, including the interesting case $D=4$. Higher gauge groups 
are needed to extend the chaotic quantization scheme to gauge fields 
in $D>5$ dimensions. Of course, this reasoning does not establish full 
chaoticity of the Yang-Mills field in these higher dimensions, it only 
indicates the possibility. Numerical studies will be required to establish 
the presence of strong chaos in these classical field theories.

The second question is more difficult. A formal answer could be 
that the Minkowski-space quantum field theory can (and even must)
be obtained by analytic continuation from the Euclidean field theory.
Any observable in the Minkowski space theory that can be expressed 
as a vacuum expectation value of field operators can be obtained in 
this manner. If this argument appears somewhat unphysical, one might 
consider a completely different approach, beginning with a chaotic 
classical field theory defined in (3+2) dimensions. Field theories 
defined in spaces with two time-like dimensions were first proposed 
by Dirac in the context of conformal field theory \cite{Dirac} and 
have recently been considered as generalizations of superstring theory 
\cite{Bars}. In this case, one time dimension is effectively eliminated
by gauge fixing. 

In the absence of similar explicit constraints, field theories with 
two timelike dimensions, even if the second time direction is compact, 
exhibit unphysical properties, such as a lack of causality and
unitarity \cite{Ynd91}. E.\ g., the Coulomb potential of a point
charge in the presence of a second, curled up timelike dimension
with period $L$ is complex:
\begin{equation}
V(r) = \frac{\alpha}{r} \left( 1 + 2\sum_{n=1}^\infty
       \E^{-2i\pi nr/L} \right) \, .
\label{eq50}
\end{equation}
The lack of causality is closely related to the problem of the 
existence of timelike closed loops, which confuse the distinction 
between past and future. These problems are avoided, if the second 
timelike dimension is ``thermal'', i.\ e. if it is compact 
in the imaginary time direction. The factor $i$ then disappears 
from the exponent of (\ref{eq50}), and the corrections to the Coulomb 
law are real and exponentially suppressed at large distances. 
In the context of the mechanism discussed in Sect.\ 5, the physical 
time dimension may be defined as the coordinate orthogonal to the 
total 5-momentum vector $P^\mu$ of the initial field configuration.
Whether this reasoning applies to the case, where the ``thermal''
field theory is really an ergodic one, remains to be confirmed, but
it is quite plausible.

In the presence of two time-like dimensions, the ``energy'' becomes a
two-component vector $\vec{E}$, which is a part of the $(D+2)$-dimensional
energy-momentum vector. If we select an initial field configuration 
with energy $E\vec{n}$, where $\vec{n}$ is a two-dimensional unit
vector, this choice defines a preferred time-like direction 
$t\vec{n}$, in which the field thermalizes. Conservation 
of the energy-momentum vector ensures that the total energy component 
orthogonal to $\vec{n}$ always remains zero. The choice of an initial 
field configuration corresponds to a spontaneous breaking of the global
SO($D$,2) symmetry down to a global SO($D$,1) symmetry. Whether this
process leads to an effective quantum field theory in the 
$(D+1)$-dimensional Minkowski space, remains to be investigated.

Finally, what about gravity? One reason, why this question is
difficult to answer, is that little is known about the properties
of general relativity as a dynamical system. Due to its different
gauge group structure, gravity has more ``capacity to resist''
chaos than the Yang-Mills fields. Local invariance against coordinate 
transformations is incompatible with the concept of Hamiltonian 
dynamics, which has a preferred time direction. The Hamiltonian
version of general relativity \cite{Misner} used in cosmology
(mixmaster universe) reflects this exceptional situation in GR.
Even the most basic definitions of deterministic chaos are not
directly applicable to general relativity and require appropriate
generalizations. Many special, chaotic solutions of Einstein's 
equations have been found \cite{NATO94,Mat00}. Most important among
these is the eternally oscillating chaotic behavior discovered in
\cite{BLK70} for the generic solution of the vacuum Einstein
equations in the vicinity of a spacelike singularity, which has the
character of deterministic chaos. However, it is not known whether 
the generic solution exhibits chaos, as in the case of the Yang-Mills 
theory. It is not even clear what the proper framework for a systematic 
numerical study of this question would be. For example, the Lyapunov 
exponents, which were so effectively used in Yang-Mills theory, are
coordinate dependent in GR due to the general covariance of Einstein's
equations against coordinate transformations.

What is clear, is that general relativity shares many of the properties, 
which allow nonabelian gauge theories to chaotically quantize themselves: 
Einstein's equations are strongly nonlinear and have a large set of gauge
invariances which could guarantee the survival of a dynamical
sector at long distances in the presence of quasi-thermal noise.
Under such conditions, GR may not even require a short-distance
cutoff, because the thermal Schwarzschild radius $(2GT)$ defines
an effective limit to short-distance dynamics which can couple
to the dynamics at large distance scales. Applying the relation
(\ref{eq48}) determining $\hbar$ for the Yang-Mills field, one
might conjecture that the analogous relation for gravity has the
form $\hbar \sim GT^2$. If the temperature parameter $T$ were of
the order of the Planck mass, this relation would yield the
observed magnitude of $\hbar$. 

However, it is not clear whether the same mechanism -- chaos,
or exponential growth of sensitivity to initial conditions --
which causes information loss in the dynamics of Yang-Mills fields,
must also operate in general relativity. 't Hooft has speculated 
that microscopic black hole formation may be the mechanism that 
causes the loss of information in the case of gravity \cite{tH99}. 
Again, a much better understanding of the structure of generic
solutions of Einstein's equations must be a prerequisite to an
exploration of these interesting questions.

\bigskip
\noindent {\em Acknowledgments:}
This work was supported in part by a grant from the
U.S. Department of Energy (DE-FG02-96ER40495), 
by the American-Hungarian Joint Fund T\'ET (JFNo. 649)
and by the Hungarian National Science Fund OTKA (T 019700).
One of us (B.M.) acknowledges the support by a 
U.~S.~Senior Scientist Award from the Alexander von Humboldt
Foundation.

\end{document}